\documentclass[dvips,twoside,fleqn]{article}
\usepackage{epsfig}
\usepackage{amstext}
\usepackage{amssymb}
\usepackage{espcrc2}
\title{The SU(2) Confining Vacuum as a Dual Superconductor}

\author{
Paolo Cea\address{Dipartimento di Fisica - Universit\`a di Bari,
               via Amendola 173, 
               70126 Bari, Italy}$^{\text{,b}}$
and Leonardo Cosmai\address{INFN - Sezione di Bari, 
               via Amendola 173, 70126 Bari, Italy}
}
       
\begin{document}

\begin{abstract}
We investigate the dual superconductivity hypothesis in pure SU(2)
lattice gauge theory.  We find evidence of the dual Meissner effect
both in the maximally Abelian gauge and without gauge fixing.
We also obtain a rather good extimation of the string tension using the
value of the London penetration length.
\end{abstract}

\maketitle

\section{INTRODUCTION}

Understanding the mechanism of quark confinement is a central problem
in the high energy physics.
According to a model conjectured long time ago by
G.~'t~Hooft~\cite{Hooft75} and S.~Mandelstam~\cite{Mandelstam76}
the confining vacuum behaves as a coherent
state of color magnetic monopoles, or, equivalently,
the vacuum resembles a magnetic (dual) superconductor. 

Up to now there is some some numerical evidence in favour of this
model~\cite{DiGiacomo96,Cea93,Haymaker93,Matsubara94,Bali95,Cea95}. 
There are also
efforts~\cite{DiGiacomo95} towards the detection of monopoles condensation in
the vacuum.

We analyzed the distribution of the color fields due to static
q$\bar{\text{q}}$ pair in SU(2) lattice gauge theory, moreover
we studied the gauge dependence of the
London penetration length by working both without gauge fixing and in
the maximally Abelian gauge~\cite{AbelianP}.
The full details of this work can be found in Ref.~\cite{Cea95}

We performed Monte Carlo simulation using the overrelaxed Metropolis
algorithm in the range  $2.45 \le \beta \le 2.7$ on lattices of sizes
$16^4$, $20^4$, and $24^4$.
After a  number of thermalization sweeps $\ge 3000$ we collect 
1 measurement every 100 upgrades in the case of SU(2) without gauge fixing,
for a total amount of 100 measurements. In order to reduce the quantum 
fluctuations we cooled the lattice configurations.
This way
quantum fluctuations are reduced by a few order of
magnitude, while the relevant physical observables survives and show
a plateau vs. cooling steps.
In the case of maximally Abelian projected SU(2) we took  
1 measurement every 50 upgrades for a total amount of 700 measurements.
Remarkably in this case no cooling is needed to get a good
signal/noise ratio.
We handled statistical errors with jackknife algorithm modified to take into
account correlations.

\section{COLOR FIELDS}

\begin{figure}[t]
\begin{center}
\epsfig{file=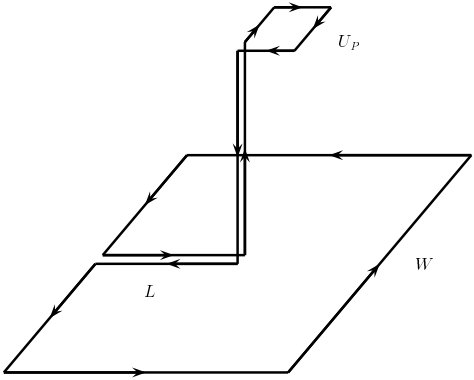,width=7.5truecm}
\end{center}
\vspace{-40pt}
\caption{The connected correlator~Eq.(\protect\ref{rhoW}) between the
plaquette $U_p$ and the Wilson loop. The subtraction appearing in the
definition of correlator is not explicitly drawn.}
\label{Fig:correlator}
\end{figure}

We can measure the color fields by means of
the correlation $\rho_W$ of a plaquette $U_p$ 
with a Wilson loop $W$~\cite{ColorF}. 
The plaquette is connected to the Wilson loop by a Schwinger line
L~(see Fig.\ref{Fig:correlator}):
\begin{equation}
\label{rhoW}
\rho_W = \frac{\left\langle \text{tr}
\left( W L U_P L^{\dagger} \right)  \right\rangle}
              { \left\langle \text{tr} (W) \right\rangle }
 - \frac{1}{2} \,
\frac{\left\langle \text{tr} (U_P) \text{tr} (W)  \right\rangle}
              { \left\langle \text{tr} (W) \right\rangle } \; .
\end{equation}
The chromoelectromagnetic field strength tensor is defined as
\begin{equation}
\label{fieldstrength}
F_{\mu\nu}(x) = \frac{\sqrt{\beta}}{2} \, \rho_W(x)   \;.
\end{equation}
\subsection{Maximally Abelian Projection}

\begin{figure}[t]
\begin{center}
\epsfig{file=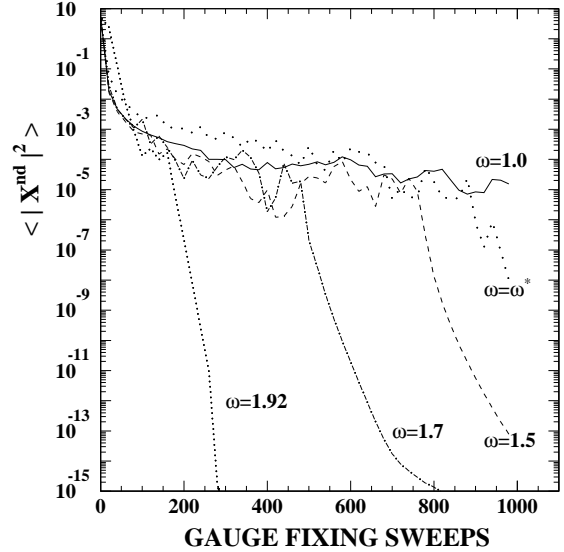,width=7.5truecm}
\end{center}
\vspace{-40pt}
\caption{Efficacy of gauge fixing defined by
Eq.~(\protect\ref{off-diag}) as a function of the overrelaxation
parameter $\omega$ for the $L=16$ lattice. The case
$\omega=\omega^\ast$ corresponds to alternate $\omega=1.0$
with $\omega=2.0$ in the gauge fixing sweeps.}
\label{Fig:overrelaxation}
\end{figure}

In the 't~Hooft formulation the dual superconductor
model is elaborated through the 
Abelian projection.
To perform the Abelian projection on the lattice we
fix the gauge by diagonalizing an operator $X(x)$ which transforms
according to the adjoint representation and then 
extract the diagonal part $U^A_\mu(x)$ out of the 
gauge transformed links 
$\widetilde{U}_\mu(x) = V(x) U_\mu(x) V^\dagger(x+\hat{\mu})$:
\begin{equation}
\label{Ua}
U^A_\mu(x) =  \text{diag} \left[ e^{i \theta^A_\mu(x)}, 
e^{-i \theta^A_\mu(x)} \right] \:,
\end{equation}
where $\theta^A_\mu(x) =  \text{arg} \left[ \widetilde{U}_\mu(x) 
\right]_{11}$.
The Abelian field strength tensor is
\begin{equation}
\label{FmunuAb}
F^A_{\mu\nu}(x) = \frac{\sqrt{\beta}}{2} \rho^A_W(x)
\end{equation}
with
\begin{equation}
\label{rhoWab}
\rho_W^{A} = \frac{\left\langle \text{tr}
\left(  W^A  U_P^A \right) \right\rangle}
              { \left\langle \text{tr} \left( W^A  \right)
\right\rangle }
 - \frac{1}{2} \,
\frac{\left\langle \text{tr}
\left( U_P^A \right) \text{tr} \left( W^A  \right) \right\rangle}
              { \left\langle \text{tr}
\left( W^A \right) \right\rangle } \; .
\end{equation}
We perform the Abelian projection in the 
maximally Abelian gauge
(in the continuum: $D_\mu A^\pm_\mu(x) = 0$). On the lattice
the gauge is fixed by maximizing 
(over all SU(2) gauge transformations $g(x)$)
the lattice functional
\begin{equation}
\label{Rl}
R_l = \sum_{x,\hat{\mu}} \frac{1}{2} \text{tr} \left[ \sigma_3 U_\mu(x)
\sigma_3 U^\dagger_\mu(x) \right] \,.
\end{equation}
This is equivalent to diagonalize ($X(x) \in$ SU(2) algebra):
\begin{eqnarray}
\label{X}
X(x) & = & \sum_\mu \left[ U_\mu(x) \sigma_3 U^\dagger_\mu(x) \right.
\nonumber \\
&  & \;\; \left. 
+ \; U^\dagger_\mu(x-\hat{\mu}) \sigma_3 U_\mu(x-\hat{\mu}) \right] \,.
\end{eqnarray}
Gauge fixing is performed by an iterative overrelaxed
algorithm~\cite{GaugeFL}.
Once we found the SU(2) matrix $g(x)$ which locally maximizes $R_l$, we keep
$g_{\text{over}}(x) = g(x)^\omega$ where the $\omega$ parameter has to be
properly tuned to increase the efficacy of the gauge 
fixing~(Fig.~\ref{Fig:overrelaxation}).
To establish a convergence criterion for the iterative gauge fixing
we consider the average size of the
non-diagonal matrix elements of $X$, Eq.(\ref{X}), over the whole lattice:
\begin{equation}
\label{off-diag}
\left\langle \left| X^{\text{nd}} \right|^2 \right\rangle =
\frac{1}{L^4} \sum_x \left[   \left| X_1 \right|^2 +  
\left| X_2 \right|^2   \right]
\end{equation}
where $X=X_1 \sigma_1 + X_1 \sigma_2 +  X_1 \sigma_3$.
The optimal overrelaxation parameter agrees with the one relevant to the Landau
gauge fixing~\cite{GaugeFL}:%
\begin{equation}
\label{omega_c}
\omega_c = \frac{2}{1+\frac{c}{L}} \,,  \quad   c \simeq 0.7 \:.
\end{equation}
\subsection{Results}
\begin{figure}[t]
\begin{center}
\epsfig{file=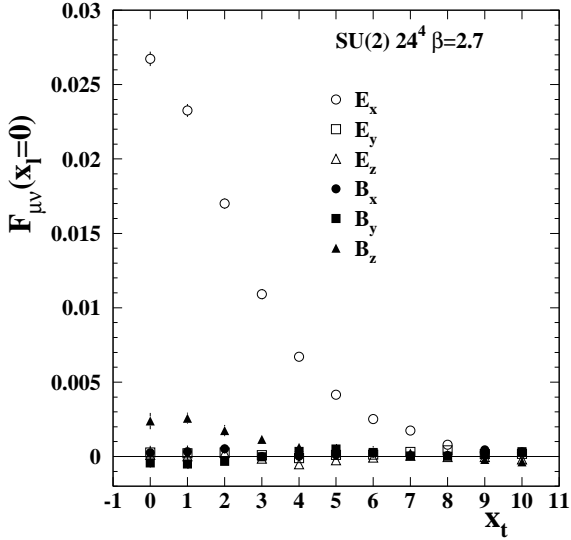,width=7.5truecm}
\end{center}
\vspace{-40pt}
\caption{The field strength tensor $F_{\mu\nu}(x_l,x_t)$ evaluated at
$x_l=0$ on a $24^4$ lattice at $\beta=2.7$, using Wilson loops of size
$10 \times 10$ in Eq.~(\protect\ref{rhoW}).}
\label{Fig:Fmunu}
\end{figure}

In both cases of SU(2) without gauge fixing and SU(2) in the maximally
Abelian gauge we obtained that (see Fig.~\ref{Fig:Fmunu} for the
case of SU(2) without gauge fixing) 
the longitudinal chromoelectric field $E_x \equiv E_l$ is sizeable
(the other field strength components are a few order of magnitude smaller).
Furthermore $E_l$ is almost constant along the $q \bar{q}$ flux tube and 
decreases fast  along the direction transverse to the flux tube.
%
%\begin{figure}[t]
%\begin{center}
%\includegraphics[width=7.5truecm]{figure_3.ps}
%\end{center}
%\vspace{-40pt}
%\caption{The maximally Abelian projected field strength tensor
%$F^A_{\mu\nu}(x_l,x_t)$ evaluated at $x_l=+1$ on a $16^4$ lattice at
%$\beta=2.5$, using Wilson loops of size $6 \times 6$ in
%Eq.~(\protect\ref{rhoWab}).}
%\label{Fig:FmunuAb}
%\end{figure}
%

\section{THE LONDON PENETRATION LENGTH}
If the dual superconductor scenario holds, the transverse shape of the
longitudinal chromoelectric field $E_l$ should resemble the dual
version of the 
Abrikosov vortex field distribution. Hence we expect
that $E_l(x_t)$ can be fitted according to 
\begin{equation}
\label{London}
E_l(x_t) = \frac{\Phi}{2 \pi} \mu^2 K_0(\mu x_t) \;, x_t > 0
\end{equation}
where $K_0$ is the modified Bessel function of order zero, $\Phi$ is
the external flux, and $\lambda=1/\mu$ is the London penetration
length. We try the fit outside the coherence region $x_t > \xi$, $\xi$ being
the coherence length, i.e. $\xi$ measures the coherence of the magnetic
monopole condensate.
We fit our data for $x_t \ge 2$ obtaining $\chi^2/f \lesssim 1$
and check the stability of the fit parameters $\mu$ and $\Phi$ by fitting
the data with the cuts $x_t \ge x_t^{\text{min}}=2,3,4,5$.
Since in the case of SU(2) without gauge fixing we adopted a cooling
procedure, we verified~\cite{Cea95} the stability of the fit 
parameter $\mu$ versus the number of cooling steps.

\begin{figure}[t]
\begin{center}
\epsfig{file=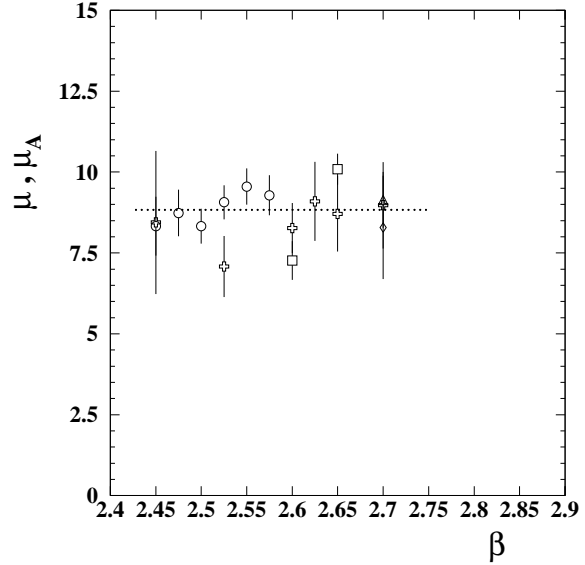,width=7.5truecm}
\end{center}
\vspace{-40pt}
\caption{$\mu$ and $\mu_A$ (in units of $\Lambda_{\overline{MS}}$)
versus $\beta$  for square Wilson loops. Circles, squares, and
triangle refer to $L=16$, $20$, $24$ respectively. Crosses and diamond
refer to the Abelian projected correlator $\rho^A_W$ with $L=16$, $20$
respectively.}
\label{Fig:mu_vs_beta_Ab-nonAb}
\end{figure}

We found that the London penetration length extracted from 
the gauge invariant correlator $\rho_W$
\begin{equation}
\frac{\mu}{\Lambda_{\overline{MS}}} = 8.96 (31) \;, 
\chi^2/f=2.11
\end{equation}
agrees with the one extracted
from the Abelian projected correlator $\rho_W^A$ 
(Fig.~\ref{Fig:mu_vs_beta_Ab-nonAb})
\begin{equation}
\frac{\mu_A}{\Lambda_{\overline{MS}}} = 8.26 (67) \;,
\chi^2/f=0.41  \,.
\end{equation}
\section{STRING TENSION}
We have shown that
the longitudinal chromoelectric field $E_l$ is almost constant along 
the flux tube (i.e. the long distance potential which feel the color
charges is linear).
We have also ascertained that the longitudinal chromoelectric 
field $E_l$ is the only sizeable
component of the field strength tensor, and its 
transverse distribution $E_l(x_t)$ can be fitted by Eq.(~\ref{London}).
Since the string tension is given by the energy stored into the 
flux tube per unit
length, using the above results and extrapolating the $K_0$-distribution up to
$x_t \rightarrow 0$ (with negligible error if $\lambda=1/\mu \gtrsim
\xi$),
%
%\begin{equation}
%\label{string_tension}
%\sigma \simeq \frac{1}{2} \int d^2 x_t E_l^2(x_l,x_t) =
%\frac{\Phi^2 \mu^2}{8 \pi} \;.
%\end{equation}
%
we get the simple relation between the string tension and  the penetration
length:
\begin{equation}
\label{sqrtstring}
\sqrt{\sigma} \simeq \frac{\Phi}{\sqrt{8 \pi}}  \mu 
\;.
\end{equation}
The main uncertainty comes out from the fit parameter $\Phi$. However we
observe that 
in the maximally Abelian projection $\Phi_A \simeq 1$,
and $\Phi \simeq \Phi_A$ when $\beta \rightarrow \infty$.
We feel that the external flux $\Phi$ is strongly affected by lattice artifacts
which, however, are strongly reduced in the maximally Abelian projection. We
can try to get rid of these effects by assuming that in the limit
$\beta\rightarrow\infty$:
$\Phi \simeq  \Phi_A \simeq  1$.
This way we get:
\begin{equation}
\label{sqrtstringdef}
\sqrt{\sigma} \simeq \frac{\mu}{\sqrt{8 \pi}}
\;.
\end{equation}
\begin{figure}[t]
\begin{center}
\epsfig{file=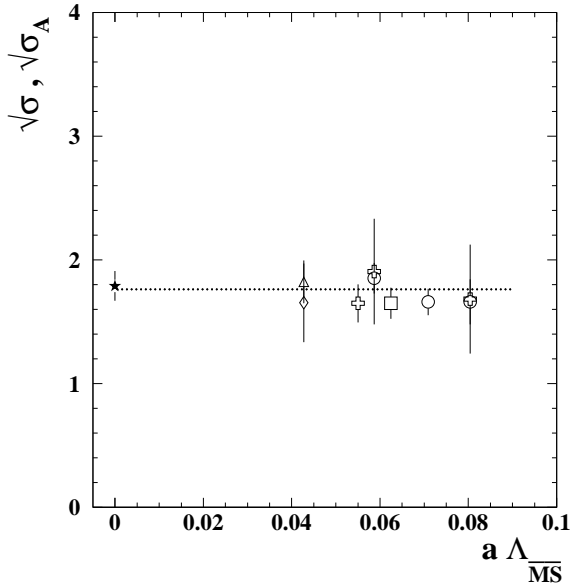,width=7.5truecm}
\end{center}
\vspace{-40pt}
\caption{String tension (in units of $\Lambda_{\overline{MS}}$)
evaluated through
Eq.~(\protect\ref{sqrtstringdef}). Star refers to the 
continuum extrapolated value of the string tension obtained using
Wilson loops on lattice larger than ours. 
Points and crosses refer to $L=16$, squares and diamond to $L=20$,
triangles to $L=24$. Crosses and diamond correspond to the maximally
Abelian gauge. For figure readability
not all the available data are displayed.}
\label{Fig:string_tension}
\end{figure}
Fitting all together the data to a constant we get (purely statistic error)
\begin{equation}
\label{stringvalue}
\frac{\sqrt{\sigma}}{\Lambda_{\overline{MS}}} = 1.76(6) \,, \qquad
\chi^2/f=1.44  \,.
\end{equation}
Our extimation of the
string tension is  consistent with the linear asymptotic extrapolation of the
string tension data extracted from Wilson loops on lattices larger
than ours~\cite{Fingberg93}
\begin{equation}
\label{wilsonstring}
\frac{\sqrt{\sigma}}{\Lambda_{\overline{MS}}} = 1.79(12) \;.
\end{equation}
Moreover note that, due to $\mu \simeq \mu_A$, we have
\begin{equation}
\label{abstreqstr}
\sqrt{\sigma} \simeq \sqrt{\sigma}_A  \:.
\end{equation}
\section{CONCLUSIONS}
We found evidence that the SU(2) vacuum behaves like a dual superconductor.
In particular, we verified that the flux tube color field is composed 
by the chromoelectric component
parallel to the line joining the static charges.
This longitudinal chromoelectric field is almost constant far from the
color sources and decreases rapidly in the direction transverse to the flux
tube.
The transverse distribution of 
the longitudinal chromoelectric field behaves according to the dual 
London equation~Eq.~(\ref{London}).
The London penetration length extracted from the Monte Carlo data
using Eq.~(\ref{London}) is a physical quantity 
$\lambda_{\text{ max. Ab. proj.}} = \lambda_{\text{SU(2)}}$.
So that the long range properties of the SU(2) confining vacuum can be described
by an effective Abelian theory.
Moreover we established a simple relation between the string
tension and  the penetration length which gives an extimate of
$\sqrt{\sigma}$ close to the extrapolated continuum limit
available in the literature.

\end{document}